\newcommand{\be}{\begin{equation}}
\newcommand{\ee}{\end{equation}}

\documentclass[aps,prl,twocolumn]{revtex4}
\usepackage{graphicx}
\usepackage{amssymb}
\usepackage{bm}

\begin{document}

\title{Self-localization of polariton condensates in periodic potentials}

\author{E. A. Ostrovskaya$^{1}$, J. Abdullaev$^{1}$, M. D. Fraser$^{2}$, A. S. Desyatnikov$^{1}$, and Yu. S. Kivshar$^{1}$}

\affiliation{$^{1}$Nonlinear Physics Centre, Research School of Physics and Engineering, The Australian National University, Canberra ACT 0200, Australia \\
$^{2}$National Institute of Informatics, 2-1-2 Hitotsubashi, Chiyoda-ku, Tokyo 101-8430, Japan}


\begin{abstract}
We predict the existence of novel spatially localized states of exciton-polariton Bose-Einstein condensates
in semiconductor microcavities with fabricated periodic in-plane potentials. Our theory shows that, under the condition of
continuous off-resonant pumping and losses associated with polariton decay,  localization is observed for a wide range of
optical pump parameters due to effective potentials self-induced by the polariton flows in the spatially periodic system.
We reveal that the localization of exciton-polaritons in the lattice may occur both in the gaps and bands of the single-particle linear
spectrum, and is dominated by the effects of gain and dissipation rather than the applied potential, in sharp contrast to the conservative condensates of ultracold alkali atoms.
\end{abstract}
\maketitle

{\em Introduction. -- } Ultracold atomic gases loaded into periodic potentials created by optical or magnetic lattices continue to provide a fascinating tool for exploring fundamental quantum many-body physics, as well as offering an opportunity to build quantum simulators and scalable quantum information processing systems \cite{qsim}. 

In recent years, polariton condensates created in semiconductor microcavities \cite{BEC06} have emerged as an attractive alternative to the atomic systems, owing to the relatively high condensation temperatures, direct momentum and real-space imaging via the cavity photoluminescence, and possibility to manipulate a condensate using both optical pump and structure of the microcavity \cite{book}.  Considerable effort has been directed towards achieving the polariton condensation in one- and two-dimensional `lattice' potentials and observing macroscopic population of the single-particle band-gap spectrum. So far, the majority of the experiments with exciton-polaritons in a periodic potential have been performed in a pulsed excitation regime, which has allowed to populate and image rapidly decaying high momentum states in a periodic potential. The staggered '$\pi$-state' phase structure of the higher-order Bloch states has been demonstrated in a 1D lattice created by metal deposition onto the microcavity \cite{stanford1d}. Similarly, condensation into second band has been achieved in a 2D square lattice \cite{stanford2d}, and the Brillouin zone structure of the Kagome lattice has been mapped out \cite{stanfordkagome}. Recently, condensation into an energy state corresponding to a single-particle spectral gap has been demonstrated in a 1D lattice created by lateral modulation of quantum wires in a microcavity \cite{gap_state}. Apart from the fabricated lattices, condensation in the tuneable acoustically created lattices has been investigated  \cite{acoustic,acoustic_tight}. 

One of the most fundamental effects due to the band-gap structure is localization of the condensate {\em in real space}. In Kagome lattices \cite{stanfordkagome} this, in principle, can be achieved by accessing the flat band of the spectrum, which signifies dispersionless dynamics. Alternatively, spatial localization in spectral gaps of the band-gap structure can be achieved due to the balance of repulsive nonlinearity and modified effective dispersion due to the negative effective mass. This mechanism has been explored in the context of gap solitons both in photonic \cite{photonic_gap} and matter-wave \cite{bec_gap} band-gap structures. Scarce theoretical descriptions of exciton-polaritons in periodic potentials  \cite{malomed_pgap,byrnes10} so far disregarded the fact that, unlike the atomic condensate in thermal equilibrium, the polariton condensate is a {\em driven, highly dissipative system} due to finite lifetime of quasiparticles and the presence of the optical pump. This distinct dissipative nature of polariton condensates becomes very important in the continuous-wave (cw) non-resonant excitation regime \cite{BEC06}, which creates long-living condensates with spontaneously established spatial coherence and long coherence times. The presence of gain-assisted localization and related dynamics may play a critical role for such dissipative condensates \cite{gain_trapping,Berloff08}, in contrast to atomic Bose-Einstein condensates (BECs) in thermal equilibrium.

In this work, we identify and analyse the mechanism for localization of a dissipative polariton condensate in a non-resonant cw excitation regime, with an external periodic potential. We show that the driven-dissipative nature of the polariton condensate gives rise to additional effective potentials local to the condensate, namely a defect-type potential introduced by the pump-induced polariton reservoir and a potential self-induced by the strongly  modulated flow of polaritons away from the excitation spot. Trapping of the polariton condensate in these effective potentials, which {\em supersede the external lattice potential}, is responsible for its spatial localization.

{\em Model. --} We assume that the near-equilibrium polariton condensate is formed via cw incoherent excitation with a spatially inhomogeneous pump. Above condensation threshold,  it can be described by the mean field open-dissipative model for a polariton condensate coupled to an incoherent polariton reservoir density \cite{Wouters07}:  
\begin{eqnarray} 
i\frac{\partial \Psi}{\partial t}&=&\left[-\frac{\nabla_\perp^2}{2}
+V_{NL}+\frac{i}{2}(Rn_R-\gamma_c)+V_L\right]\Psi,  \nonumber \\  
\frac{\partial n_R}{\partial t}&=&-(\gamma_R+R|\Psi|^2)n_R(\vec{r},t)+P(\vec{r}). \label{model} 
\end{eqnarray}
Here $\Psi$ is the condensate wavefunction, $V_{L}(\vec{r})$ is a linear periodic potential created for the condensate, and  $V_{NL}=g_c|\Psi|^2+g_Rn_R$ is an effective nonlinear potential depending both on the condensate density and the reservoir density, $n_R$, which is determined by the inhomogeneous optical pump $P(\vec{r})$. The parameters defining the condensate dynamics are the relaxation rates of the condensed polaritons $\gamma_{c}$ and reservoir polaritons $\gamma_{R}$, the stimulated scattering rate $R$, and the polariton-polariton interaction strengths $g_R$, $g_c$. We have written the model (\ref{model}) in a dimensionless form by using the characteristic scaling units of time $1/\gamma_c$, energy $\hbar \gamma_c$, and length $[\hbar/(m_{LP}\gamma_c)]^{1/2}$, where $m_{LP}$ is the lower-polariton effective mass \cite{dissipative_pra}. Under this scaling $\gamma_c=1$, however we formally retain this parameter hereafter. The scaling units are calculated using the parameters close to those of the experimental setup of \cite{Vortex11}, with $m_{LP}=10^{-3}m_e$, $g_c=6\times 10^{-3}$ $meV$$\mu m^2$, $g_R=2g_c$, $\gamma_c=0.02$ $ps^{-1}$, $\gamma_R=1.5 \gamma_c$, and $R=0.01$ $\mu m^2$ $ps^{-1}$.

\begin{figure}
\includegraphics[width=8.5 cm]{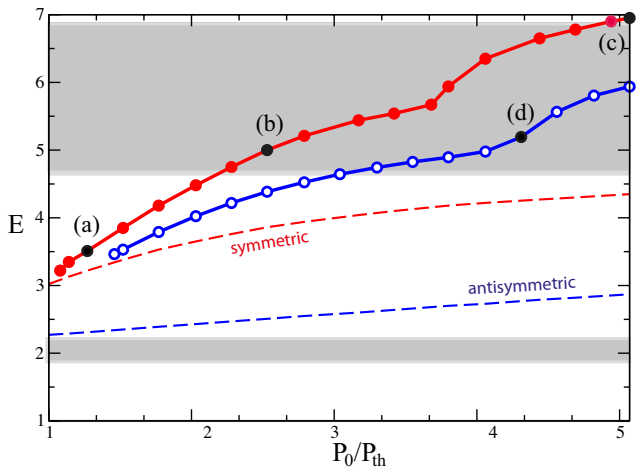} 
\caption{Steady state energies of the polariton BEC obtained with cw on-site (closed circles) or $\pi/4$ off-site (open circles) narrow Gaussian pump with the width $\sigma^2=0.6$. Shaded (open) areas show bands (gaps) of the single-particle Bloch-wave spectrum in the 1D
 lattice potential $V_L(x)$ ($V_0=5.0$, $k_l=1.5$).  Dashed lines are linear defect states in the combined effective potential $V_L(x)+g_R P(x)/\gamma_R$ (see text).} \label{fig1}
\end{figure}

\begin{figure}
\includegraphics[width=8.5 cm]{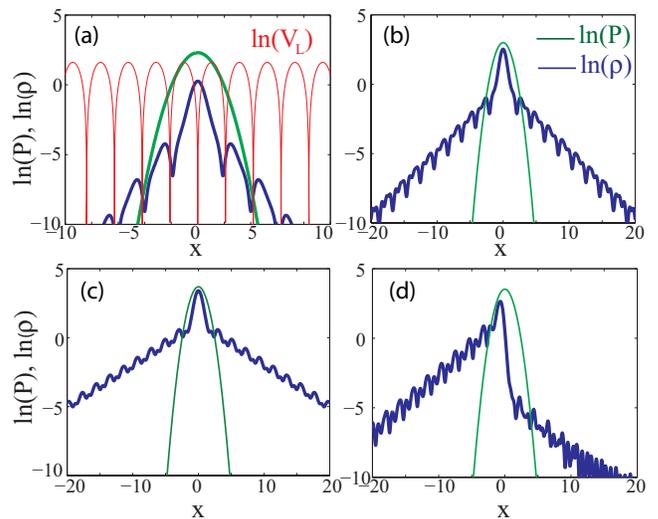} 
\caption{Spatial profiles of the condensate density $\ln(\rho(x))$ and pump intensity $\ln(P(x))$ in a periodic 1D lattice potential $V_L(x)$ with $V_0=5.0$ and $k_l=1.5$, corresponding to the peak pump intensity $P_0/P_{th}=1.27$ [point (a) in Fig. 1], $P_0/P_{th}=2.53$ [point (b)], $P_0/P_{th}=5.06$  [point (c)] , and $P_0/P_{th}=4.30$ [point (d)]. In addition, panel (a) shows the lattice potential $\ln(V_L)$ (note the difference in the position scale).} \label{fig2}
\end{figure}

{\em localization in 1D. --} We begin our consideration with a simplest, one-dimensional (1D) model of a quasi-1D polariton condensate (e.g., confined in a microwire \cite{bloch_wires}) subject to a periodic potential approximated by a harmonic function $ V_{\rm L}(x) = V_0 \sin^{2}(k_l x)$, with the lattice constant $a=\pi/k_l$.  The periodicity of the potential imposes a band-gap structure onto the LP branch of the polariton dispersion curve \cite{stanford1d}, and we can consider population of this band-gap structure by polaritons depending on the pump intensity and width. To this end, we look for the steady state solutions $\Psi(x)=\psi(x)\exp(iEt)$ of the model (\ref{model}) governed by the stationary complex Ginzburg-Landau equation with inhomogeneous nonlinear gain and linear loss.
We will consider the inhomogeneous Gaussian pump profile $P(x)=P_0\exp(-\sigma^2x^2)$ and relate the peak intensity of the pump to the threshold homogeneous pump intensity $P_{th}=\gamma_R\gamma_c/R$ \cite{Wouters07}.

As has been established both experimentally (see, e.g., \cite{gain_trapping,bloch_wires}) and theoretically \cite{wouters08,Berloff08,dissipative_pra}, cw non-resonant pump leads to spatial localization of polariton BEC even without a trapping potential. This effect, resulting in formation of {\em dissipative solitons} \cite{dissipative_pra,dissipative_solitons} is due to the balance of gain and loss in the system and is irrespective of the sign of the nonlinear interactions. This mechanism is dramatically different from conventional balance of nonlinearity and dispersion (diffraction) responsible for formation of bright solitons in the coherent fields. The repulsive polariton-polariton interaction then leads to considerable spatial extension of the polariton condensate beyond the excitation spot \cite{gain_trapping,bloch_wires,Berloff08,dissipative_pra}, and results in density deformations which are especially pronounced in the presence of a trap \cite{Balili07,Berloff08,dissipative_pra}. In what follows, we will demonstrate that spatial localization of the condensate due to the balance between gain and loss persists in the presence of a periodic potential.

\begin{figure}
\includegraphics[width=8.5 cm]{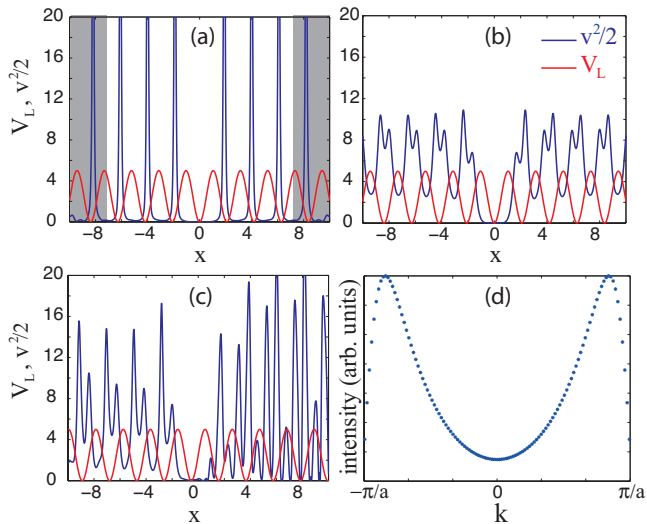} 
\caption{Spatial profiles of the lattice potential $V_L(x)$ ($V_0=5.0$, $k_l=1.5$)  and the effective linear potential due to the phase gradient $v^2/2$ corresponding to the peak on-site pump intensities (a) $P_0/P_{th}=1.27$,  (b) $P_0/P_{th}=5.06$, and (c) a non-centered pump ($\pi/4$ off-site) with $P_0/P_{th}=4.30$. Panel (d) shows the $k$-space spectrum of the steady state [(a) in Fig. \ref{fig2}] corresponding to the effective potential in (a). } \label{fig3}
\end{figure}

{\em Narrow excitation. --} A spatially narrow excitation, $P(x)$, can selectively target a single lattice site and therefore we can distinguish between an on-site and off-site excitation. In the later case, the peak of the excitation spot can be arbitrarily offset from the minimum of the lattice potential. As can be seen from examining Eq. (\ref{model}), in the case of the pump intensity near the threshold for the condensate formation, $P_0/P_{th}\sim1$, nonlinear effects can be neglected. In this case, assuming a straightforward balance of gain and loss terms, Eq.  (\ref{model}) would describe a condensate in a linear periodic potential $V_L(x)$ with the addition of a linear defect induced by the pump, $g_Rn^0_R\approx (g_R/\gamma_R)P(x)$, where $n^0_R=P/(\gamma_R+R|\psi|^2)$ is the steady state reservoir density. The energy of the lowest order linear defect modes is plotted as a function of pump intensity in Fig. \ref{fig1}, together with the band-gap spectrum imposed by the linear periodic potential. 

Taking this spectrum as a frame of reference, we can examine the energy and spatial structure of the localized steady state depending on the pump intensity. The respective dependence is plotted in Fig. \ref{fig1} and shows that, near the threshold on-site excitation, the energy of the localized state is indeed very close to that of a symmetric linear defect mode. As the ratio $P_0/P_{th}$ grows, so does the energy of the localized state until it crosses into the band of the linear band-gap spectrum. This is in contrast to dissipative photonic lattice solitons \cite{malomed_dissipative} and defect states \cite{kartashov}, which have been previously found to exist exclusively in the gaps of the band-gap spectrum. An additional feature of the polariton steady states is transition, with growing pump intensity, from narrow states localized in a vicinity of a single potential well [Fig. \ref{fig2}(a)] to broader states with exponentially decaying and spatially modulated tails [Fig. \ref{fig2}(b,c)]. 

To understand the spatial structure of the localized states, we perform the standard Madelung transformation $\psi(x)= \sqrt{\rho(x)}\exp[i\phi(x)]$, and write down the steady state equations for the condensate density $\Phi=\sqrt{\rho}$ and phase gradient (flow velocity) $v=\phi_x$:
\begin{eqnarray}
\Phi_{xx}- 2\left(V_L+g_c\Phi^2+g_Rn^0_R+\frac{v^2}{2}\right)\Phi+2E\Phi=0,\\
(\Phi^2v)_x-(Rn^0_R-\gamma_c)\Phi^2=0, \label{fluid} \nonumber
\end{eqnarray}
Given the rapid decay of the inhomogeneous pump at large $x$, we can immediately see that the asymptotic behaviour of the condensate near-linear density away from the pump spot is determined by the combination of the linear periodic potential $V_L(x)$ and a {\em self-induced} effective potential due to the phase gradient, $v^2(x)/2$. These two potentials for the narrow and wide states (a) and (b,c) in Figs. \ref{fig1},\ref{fig2} are shown in Fig. \ref{fig3}(a-c). The strong contribution of the Bloch states from the edges of the Brillouin zone  into the composition of the localized state [Fig. \ref{fig3}(d)] leads to steep spatial gradients of the phase, $\phi(x)$. As a result, the flow of the condensate out of the excitation spot induces a periodic, tight-binding Kronig-Penney-like potential [Fig. \ref{fig3} (a)], and the band-gap spectrum due to the linear lattice potential becomes of {\em little relevance} to the energy eigenstates. The sharp peaks of the effective potential are offset from the maxima of the lattice potential, which leads to the fact that the condensate density peaks can coincide with the {\em maxima} of the lattice potential [see Fig. \ref{fig2}(a)]. The tunnelling of the condensate out of the excitation spot is strongly inhibited by the tight-binding effective potential, until the modulations of the flow are smoothed out by the stronger pump. As demonstrated in Figs. \ref{fig2}(b,c) the strongly pumped condensate density then extends over many wells of the periodic potential. We stress that, in contrast to the lattice potential which exists regardless of the condensate formation, the self-induced potential can become physically meaningless far on the tails of the condensate [shaded areas in Fig. \ref{fig3} (a)], where the density is negligible and any dynamical fluctuations can quickly destroy regular phase structure.

Although the narrow gap states [Fig. \ref{fig2}(a)] greatly resemble well localized gap solitons of a conservative BEC \cite{bec_gap}, they have no connection to the Bloch states at the band edge of the first Brillouin zone in the linear lattice potential, rather branching off the linear defect states. Their spatial extent is determined purely by the phase gradients established due to the balance of gain and loss in the system. The energy of the localized state can lie in either a spectral gap or band of the lattice potential since it is supported by the effective potentials created by the polariton flow.

In the off-site excitation regime, the localized states can also be formed both within gaps and bands of the linear band-gap spectrum of the potential $V_L(x)$ [Fig. \ref{fig1}], and the threshold for the formation of these states is higher than that of the on-site state. Remarkably, the asymmetry of the defect created by the offset pump translates to the asymmetry of both the effective potential generated by the polariton flow and the condensate density [Fig. \ref{fig2}(d) and Fig. \ref{fig3}(c)]. Such asymmetric localized states are supported by the interface between two different effective potentials, and therefore resemble `mixed-gap' interface states of an atomic BEC in anharmonic lattices \cite{anharmonic}.

\begin{figure}
\includegraphics[width=8.5 cm]{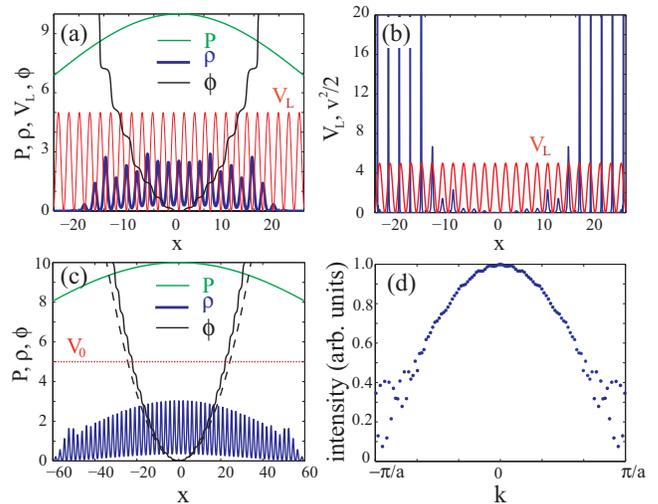} 
\caption{(a,c) Spatial profiles of the condensate density $\rho(x)$, phase $\phi(x)$, and lattice potential $V_L(x)$ ($V_0=5.0$, $k_l=1.5$)  corresponding to the peak on-site pump intensity $P_0/P_{th}=1.27$ and widths $\sigma^2=6\times 10^{-4}$ and $\sigma^2=6\times 10^{-5}$, respectively. Dashed curve in (c) is the analytical approximation of $\phi(x)$, Eq. (\ref{ho}). (b) Lattice potential $V_L$ and the self-induced potential, $v^2/2$, corresponding to (a). (d) Spectrum of the state (c) within the first Brillouin zone of the lattice potential.} \label{fig4}
\end{figure}

{\em Wide excitation. --} By fixing the pump intensity and extending the width of the pump spot significantly beyond the single lattice well, we find a variety of spatially extended nearly energy-degenerate localized states. One example of such a gap state and its supporting effective potential is presented in Figs. \ref{fig4}(a,b). These states strongly resemble `flat gap states' formed by the nonlinearity-induced truncation of nonlinear Bloch states in a conservative condensate \cite{truncated}. However, the localization in this case is due to the effective potential created by the polariton flow out of the finite gain region.

A very wide, locally homogeneous pump near condensation threshold generates extended states that locally resemble a ground state Bloch wave [Fig. \ref{fig4} (c)] of the linear lattice potential $V_L(x)$. In this regime, the asymptotic approximation of the phase profile at $x\to 0$, derived in \cite{dissipative_pra} for the case of untrapped polariton condensate becomes valid for the most of the excitation region [Fig. \ref{fig4} (c), dashed line], and the effective potential due to the polariton flow becomes harmonic:
\begin{equation}
v^2(x)\approx\frac{\gamma^2_c}{4}\left(\frac{P_0}{P_{th}}-1\right)^2x^2.
\label{ho}
\end{equation}
The density profile is a typical ground state supported by the combination of the periodic and harmonic effective potential [Fig. \ref{fig4} (c)], and its $k$-space spectrum [Fig. \ref{fig4} (d)] displays localization around $k=0$ characteristic of the ground state in the lattice \cite{stanford1d}. Thus, in this regime the dissipative condensate in a periodic potential mimics the behaviour of a conservative condensate in the {\em combined} periodic and harmonic potentials.

\begin{figure}
\includegraphics[width=8.5 cm]{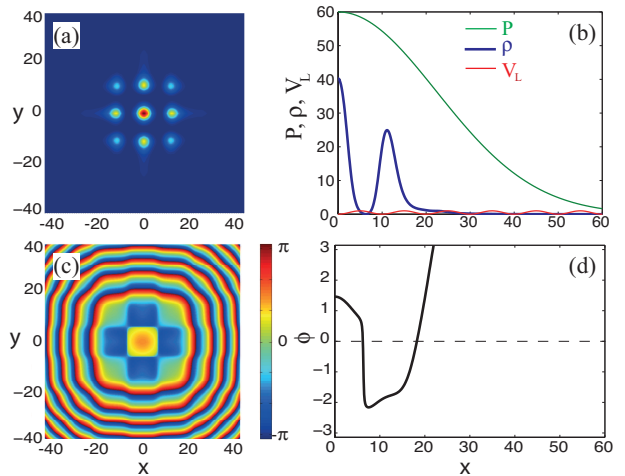} 
\caption{Spatial profiles and cross-sections of the (a,b) condensate density and (c,d) phase,  corresponding to a narrow in-gap state in a 2D lattice ($V_0=2.0$, $k_l=\pi/4$).} \label{fig5}
\end{figure}

{\em Localization in 2D. --} Similarly to the case of 1D periodic potential, we find that excitation spots of varying width can create narrow and wide localized steady states of the polariton condensate both in the gaps and bands of the single-particle band-gap spectrum corresponding to a 2D `square' lattice potential $V_L=V_0[\sin^2(k_lx)+\sin^2(k_ly)]$. Examples of a narrow state density and its phase are shown in Fig. \ref{fig5} for the case of a weak lattice. Although the detailed description of such states is beyond the scope of this work, we note that the phase structure displays sharp domain walls associated with strong phase gradients, similar to those in 1D case, which leads to the tight spatial localization of the condensate.

{\em Conclusions. --} We have demonstrated that, under the conditions of spatially inhomogeneous non-resonant cw excitation, a polariton condensate can be spatially localized in a periodic in-plane potential owing to the polariton flows established due to the balance between gain and loss. Periodic potential strongly modulates the velocity of the polariton flow out of the pumping region, and the polaritons are self-trapped in the resulting effective potential. This mechanism of localization ensures that, depending on the pump intensity, condensate can be formed and localized both in gaps and bands of the linear single-particle band-gap spectrum of the periodic potential. Localization of the dissipative condensate is therefore irrelevant to the requirement of the negative effective mass which is essential for the gap soliton formation in conservative (atomic) BECs with repulsive interparticle interactions \cite{bec_gap}.

The predicted features of the real-space localization of a dissipative polariton condensate subject to a periodic potential have strong consequences for any proposed polariton-based devices incorporating modulated potentials, and should be investigated further, both theoretically and experimentaly.  Polariton condensates in `lattice' traps, created in the non-resonant cw excitation regime explored here, are ideally suited to such studies.

This work was supported by the Australian Research Council. Discussions with D. Tanese, G. Malpuech, S. H\"offing, and A. Kavokin are gratefully acknowledged.

\end{document}